\documentclass[times,preprint]{elsarticle}

\usepackage{jcomp}
\usepackage{framed,multirow}

\usepackage{amssymb}
\usepackage{latexsym}
\usepackage{bm}
 \usepackage{indentfirst}
\usepackage{amsmath,amsthm,amssymb}

\usepackage{graphicx}
\usepackage{subcaption}
\usepackage{float}
\usepackage[framemethod=tikz]{mdframed}
\usepackage{epstopdf}

\usepackage{import}

\newcommand{\velx}{u}

\newif\ifturb

\newif\ifsutherland

\newcommand{\mask}{\chi}
\newcommand{\IBMparam}{\eta}

\newcommand{\advcoef}{c}

\usepackage{url}
\usepackage{xcolor}
\definecolor{newcolor}{rgb}{.8,.349,.1}

\journal{Journal of Computational Physics}

\begin{document}

\verso{Jiaqing Kou \textit{et. al}}

\begin{frontmatter}

\title{A combined volume penalization / selective frequency damping approach for immersed boundary methods: application to moving geometries}

\author[aachen]{Jiaqing \snm{Kou}}
\cortext[cor1]{Corresponding author}
\ead{jiaqingkou@gmail.com}
\author[upm,ccs]{Esteban \snm{Ferrer}\corref{cor1}}
\ead{esteban.ferrer@upm.es}

\address[aachen]{Institute of Aerodynamics, RWTH Aachen University, Wüllnerstraße 5a, 52062 Aachen, Germany}
\address[upm]{ETSIAE-UPM-School of Aeronautics, Universidad Politécnica de Madrid, Plaza Cardenal Cisneros 3, E-28040 Madrid, Spain}
\address[ccs]{Center for Computational Simulation, Universidad Politécnica de Madrid, Campus de Montegancedo, Boadilla del Monte, 28660 Madrid, Spain}


\begin{abstract}
\end{abstract}


\end{frontmatter}


\section{Introduction}
In recent years, high-order methods such as Flux Reconstruction (FR) and Discontinuous Galerkin (DG) have gained popularity in the field of fluid dynamics. These methods offer several advantageous numerical properties, including enhanced accuracy and reduced dissipation/dispersion errors, particularly at low to medium wavenumbers \cite{wang2013high}. Despite the advantages a well-known bottleneck of high-order methods is the necessity of high-order curved grids near body-fitted boundaries, which can complicate the process of generating the mesh. To facilitate the generation of meshes, Immersed Boundary Method (IBM) has been proposed \citep{mittal2005immersed,sotiropoulos2014immersed,griffith2020immersed}, to avoid body-fitted meshes while accurately resolving complex flows using simple grids. 

\begin{figure*}[htbp]
	\begin{subfigure}{.4\textwidth}
 		\centering
		\includegraphics[width=0.7\textwidth]{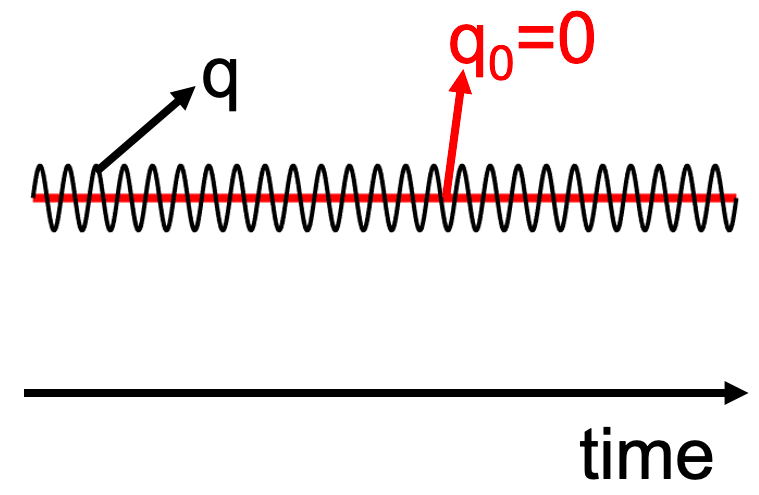}
		\caption{}
	\end{subfigure}
	\begin{subfigure}{.4\textwidth}
 		\centering
		\includegraphics[width=0.7\textwidth]{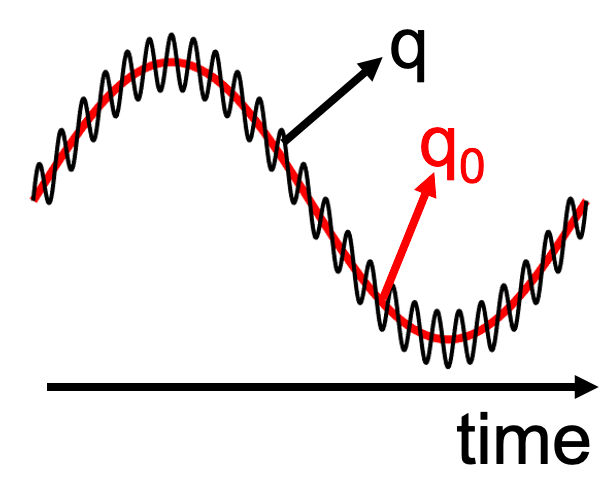}
		\caption{}
	\end{subfigure}
	\centering
	\caption{Schematic illustrations of selective frequency damping for static and moving boundary problems: a) Static geometry with zero velocity $\boldsymbol{q}_0=\boldsymbol0$ (red); b) Moving geometry with velocity $\boldsymbol{q}_0\neq\boldsymbol0$ (red).}
	\label{fig:SFD-cartoon}
\end{figure*}
Combinations of high-order methods and IBM have been proposed to merge the numerical advantages of high-order methods (spectral convergence, low dissipation/dispersion errors, etc.) with ease for mesh generation. For example, Fidkowski and Darmofal \citep{fidkowski2007triangular} and M{\"u}ller et al. \citep{muller2017high} applied cut-cell method to simulate steady compressible flows based on discontinuous Galerkin. In contrast, our recent work has shown that IBM based on Volume Penalization (VP) is a robust and easy-to-implement method for high-order discretizations (e.g., FR or DG) \cite{kou2021IBMFR2}. Through eigensolution analyses \citep{kou2021VonNeumann}, the error characteristics of this approach were clarified, showing that the penalization imposes the boundary condition by adding additional dissipation inside the solid. These works motivated our recent extension of an alternative IBM treatment combining VP and selective frequency damping (SFD) \citep{kou2022combined}, which was tested using high-order methods and static geometries. The VP+SFD approach offers additional damping inside the solid that improves the overall accuracy of the scheme. 

The purpose of this note is to extend the combined method: VP+SFD, proposed in \citep{kou2022combined} to moving geometries. This extension requires several additions: 1) Reformulate the SFD formulation for moving boundaries; 2) New manufactured solutions to test numerical convergence; 3) Numerical validation for the Navier-Stokes equations. 

\section{Methodology}
Volume penalization \cite{angot1999penalization,schneider2015immersed} imposes boundary conditions by introducing penalizing source terms inside the solid. A mask function $\mask(\boldsymbol{x},t)$ is defined to distinguish between the fluid region $\Omega _{f}$ and solid region $\Omega _{s}$:

\begin{equation}
\mask  (\boldsymbol{x},t) = \left\{\begin{matrix}1, \, \, \text{if}\, \, \boldsymbol{x} \in \Omega _{s}
\\0,\, \, \text{otherwise}
\end{matrix}\right. .
\end{equation}

For a given fluid system, written in a semi-discrete form, the VP (for Dirichlet boundary conditions) can be formulated as $\frac{d \bm{U}}{d t} = \boldsymbol{RHS} (\bm{U}) + \frac{\mask}{\IBMparam} (\bm{U}_s-\bm{U})$, where $\bm{U}$ denotes variable vector (e.g., conservative variables in the Navier-Stokes system) and $\boldsymbol{RHS}$ gathers all terms including spatial derivatives and sources. The rightmost term comes from VP formulation, where $\IBMparam$ is the penalization parameter that controls accuracy (the modeling error decays when $\eta \rightarrow 0$ \citep{angot1999penalization}). Here we discretize $\boldsymbol{RHS}$ using a high-order flux reconstruction (FR) scheme, but other schemes could also be used. Further details on the governing equations, the FR discretization, and error analysis are given in \cite{kou2021IBMFR2,kou2021VonNeumann,kou2022combined}.



\begin{figure*}[htbp]
	\begin{subfigure}{.48\textwidth}
 		\centering
		\includegraphics[width=0.7\textwidth]{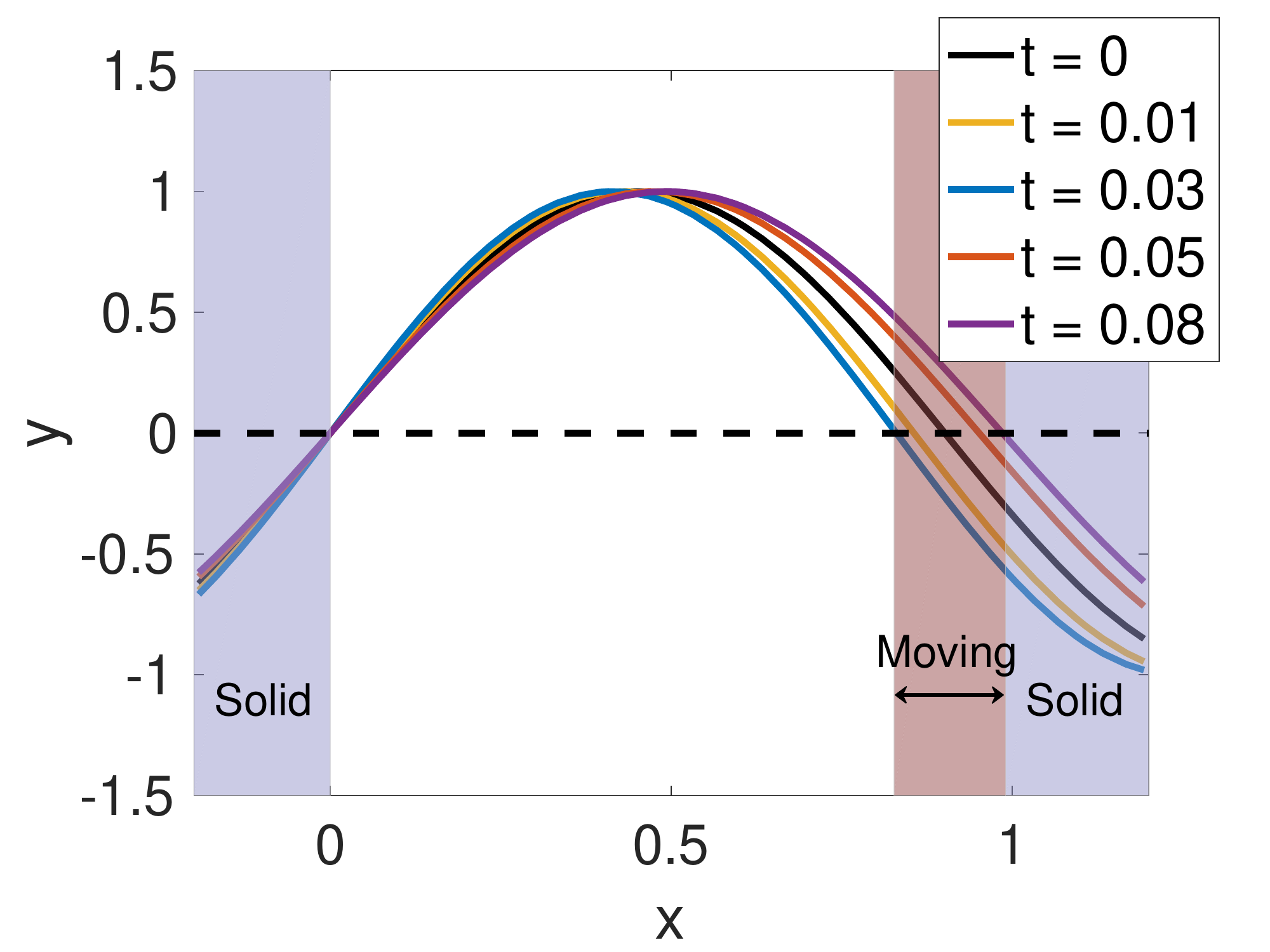}
		\caption{}
	\end{subfigure}
	\begin{subfigure}{.48\textwidth}
 		\centering
		\includegraphics[width=0.7\textwidth]{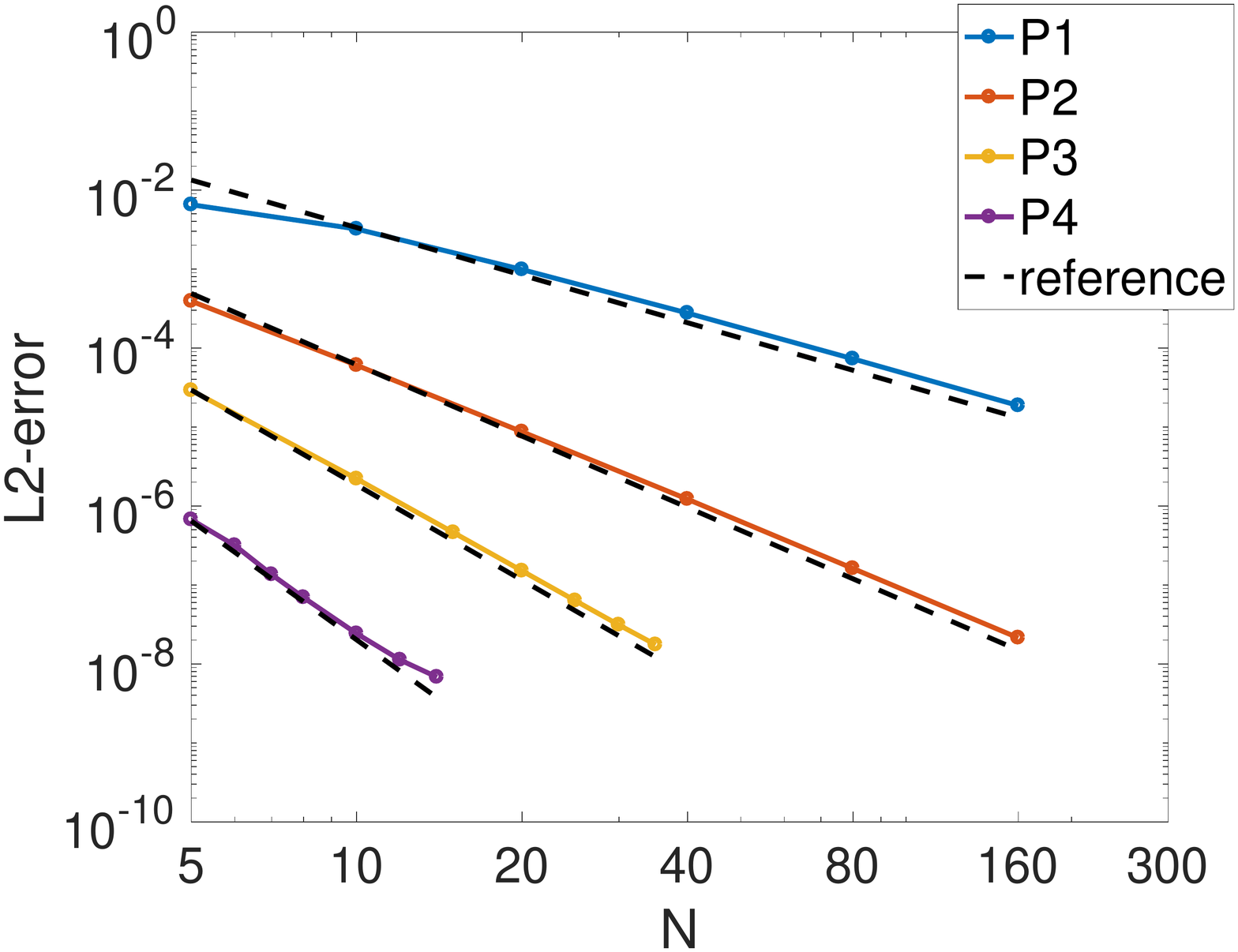}
		\caption{}
	\end{subfigure}
	\centering
	\caption{Method of manufactured solutions: a) Schematic illustration; b) Convergence in the flow region.}
	\label{fig:MMS-cartoon}
\end{figure*}

The SFD method was proposed by Åkervik et al. \citep{aakervik2006steady} to compute unstable steady solutions (unstable fixed points) in global stability analyses, and enforces steady states by damping unstable modes (with characteristic frequencies). The method is motivated by control theory where a proportional feedback control is introduced as a forcing term to drive the solution $\boldsymbol{q}$ towards the target solution. For moving bodies, the numerical solution $\boldsymbol{q}$ can be decomposed into the moving value of the geometry $\boldsymbol{q}_0$ and an oscillatory part $\boldsymbol{q}'$ (representing spurious waves), see figure \ref{fig:SFD-cartoon}: $\boldsymbol{q} = \boldsymbol{q}_0 + \boldsymbol{q}'$. Note that the former term is zero for fixed boundaries. The oscillatory part $\boldsymbol{q}'$ refers to the non-physical (spurious), high-frequency scales that need to be damped by the SFD approach. It should be noted that the motion frequency is assumed to be smaller than the frequency of the spurious oscillations, and we assume enough spectral separation between $\boldsymbol{q}_0$ and $\boldsymbol{q}'$. Under these assumptions, the SFD formulation can be reformulated as:

\begin{equation}
    \left\{\begin{matrix}
    \dot{\boldsymbol{q}} = \boldsymbol{f}(\boldsymbol{q}) - \chi_{f} (\boldsymbol{q} - \boldsymbol{q}_0 - \bar{\boldsymbol{q}})
\\  \dot{\bar{\boldsymbol{q}}} = (\boldsymbol{q} - \boldsymbol{q}_0 -  \bar{\boldsymbol{q}}) / \Delta
\end{matrix}\right. ,
\end{equation}
where $\chi_f$ is the control coefficient, and $\boldsymbol{f}$ is the nonlinear operator applied to a state variable vector $\boldsymbol{q}$ (i.e., the right-hand-side term). $\Delta$ is the filter width to control the cutoff frequency $\omega_c = 1 / \Delta$. The target low-pass filtered solution $\bar{\boldsymbol{q}}$ is time-varying and this approach drives $\boldsymbol{q}$ to the target solution. When this approach is used for immersed boundary treatment, the SFD term is only applied inside the solid region, resulting in minimum computation overhead.

To implement this approach to a Navier-Stokes flow solver with minimal effort, two splitting schemes for the VP and the SFD are used, as detailed in \cite{kou2022combined}. Using eigensolution analyses \cite{kou2021VonNeumann}, we proposed guidelines to select the key parameters in the combined method: the penalization parameter is fixed to $\IBMparam = \Delta t$, the control parameter of the SFD should be $\chi_f = 1/\IBMparam$ while the filter width $\Delta$ needs to be large enough to remove all spurious oscillations of the numerical solution (e.g., $\Delta > 10$). Finally, note that the VP and SFD are only applied inside the geometry (to mimic the effect of solid boundaries), while no treatment is applied to the fluid region.

\section{Numerical Validation}
The proposed method is validated by two test cases, where FR is used for spatial discretization. The first one shows the numerical convergence, and the second one shows the validity of the approach for moving objects.

\begin{figure*}[htbp]
	\centering
		\includegraphics[width=0.4\textwidth]{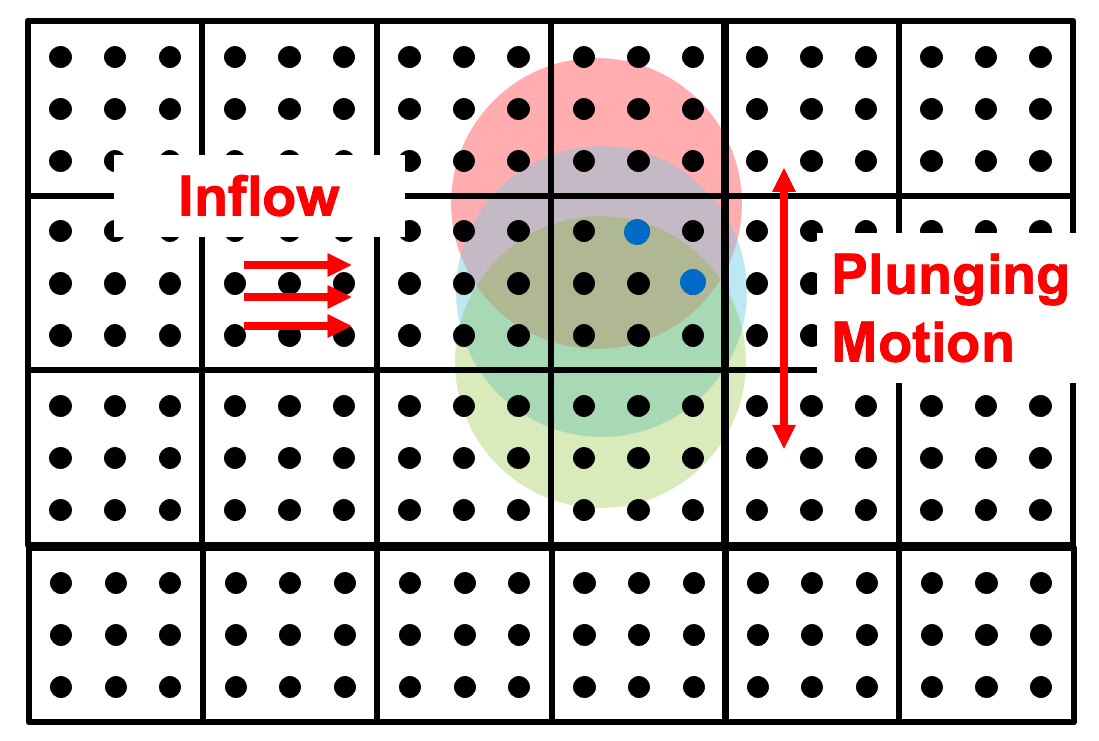}
	\caption{Moving cylinder: Schematic illustration of flow over a moving cylinder. Selected Gauss points for probing the solution are depicted in blue. }
	\label{fig:Cyl-cartoon}
\end{figure*}

\subsection{Method of manufactured solutions for moving boundaries}
The method of manufactured solution is commonly used for code verification. To test our IBM implementation, we modify the case proposed by Tremblay et al. \cite{tremblay2006code}, which is originally used to evaluate the error in fluid-structure interaction problems. Considering the one-dimensional advection-diffusion problem:

\begin{equation}
    \frac{\partial \velx}{\partial t} + \advcoef \frac{\partial \velx}{\partial x} - \nu \frac{\partial^2 u}{\partial x^2} + \frac{\mask}{\IBMparam}(\velx - \velx_s(x,t)) = 0.
    \label{eq:adv}
\end{equation}

Note that the penalized solution $\velx_s(x,t)$ and the mask function $\mask(x,t)$ vary in space and time. We consider the moving physical domain $x \in [0, 1/L(t)]$, where the right interface parametrized by $L$ is oscillating in time, to mimic the moving boundary. We first design the analytical solution and the flow domain, with user-defined parameters $A_m$, $B_m$, $C_m$ and $D_m$: $U(x,t) = A_m + \text{sin}[\pi L(t) x], \ \  L(t) = B_m+C_m\text{sin}(D_m \pi t)$. The one-dimensional advection-diffusion equation, without the source term, can be written into a linear operator: $\mathcal{L}(u) = u_t + cu_x - \nu u_{xx}$. In the method of manufactured solution, the source term $Q$ is determined to produce the solution $U$, and obtained by applying the operator $\mathcal{L}$ to $U$ as: $   Q(x,t) = \mathcal{L}(U) = \ C_m D_m \pi^2 x \cdot \text{cos}[\pi L(t)x] \cdot \text{cos}(\pi D_m t) + c \pi L(t) \cdot \text{cos}[\pi L(t) x]  \nonumber + \nu [\pi L(t)]^2 \cdot \text{sin}[\pi L(t) x ]  \nonumber $. By adding this term to the right-hand side of the governing equation, the numerical error can be obtained by comparing the residual against zero. Finally, when considering the immersed boundary, we obtain the modified equation for the moving geometry: $u_t + c u_x - \nu u_{xx} = (1-\mask)Q(x,t) + S$, where $S = -\frac{\mask}{\IBMparam}(\velx - \velx_s(x,t))$ is the VP term and the source term $Q(x,t)$ is only added inside the fluid region. The time-dependent mask varies in time to maintain Dirichlet boundary conditions $u=0$ at interfaces, which is one when $x \leq 0$ or $x \geq 1/L(t) $ and zero otherwise.
%
%


\begin{figure*}[htbp]
	\begin{subfigure}{.48\textwidth}
 		\centering
		\includegraphics[width=0.7\textwidth]{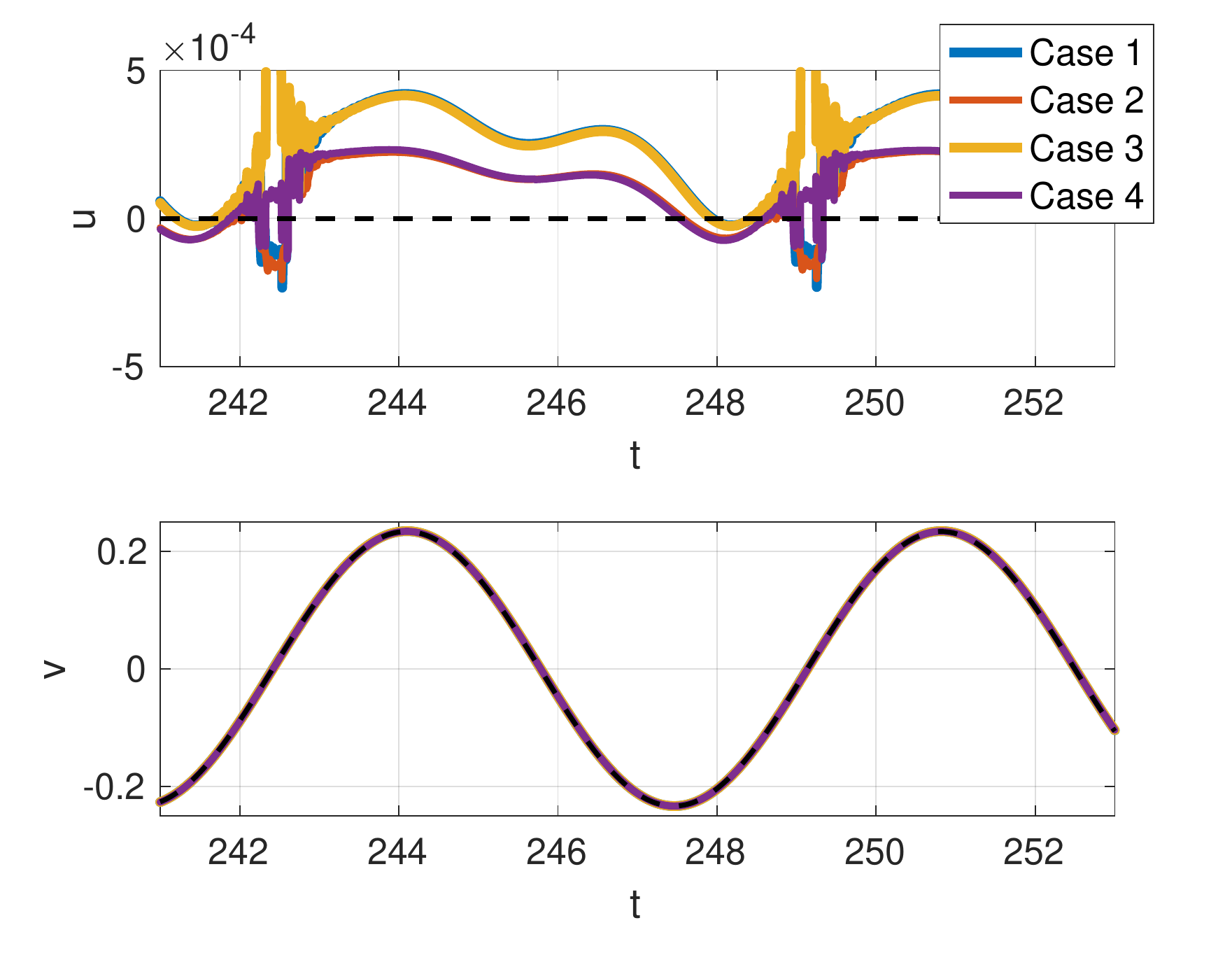}
		\caption{}
	\end{subfigure}
	\begin{subfigure}{.48\textwidth}
 		\centering
		\includegraphics[width=0.7\textwidth]{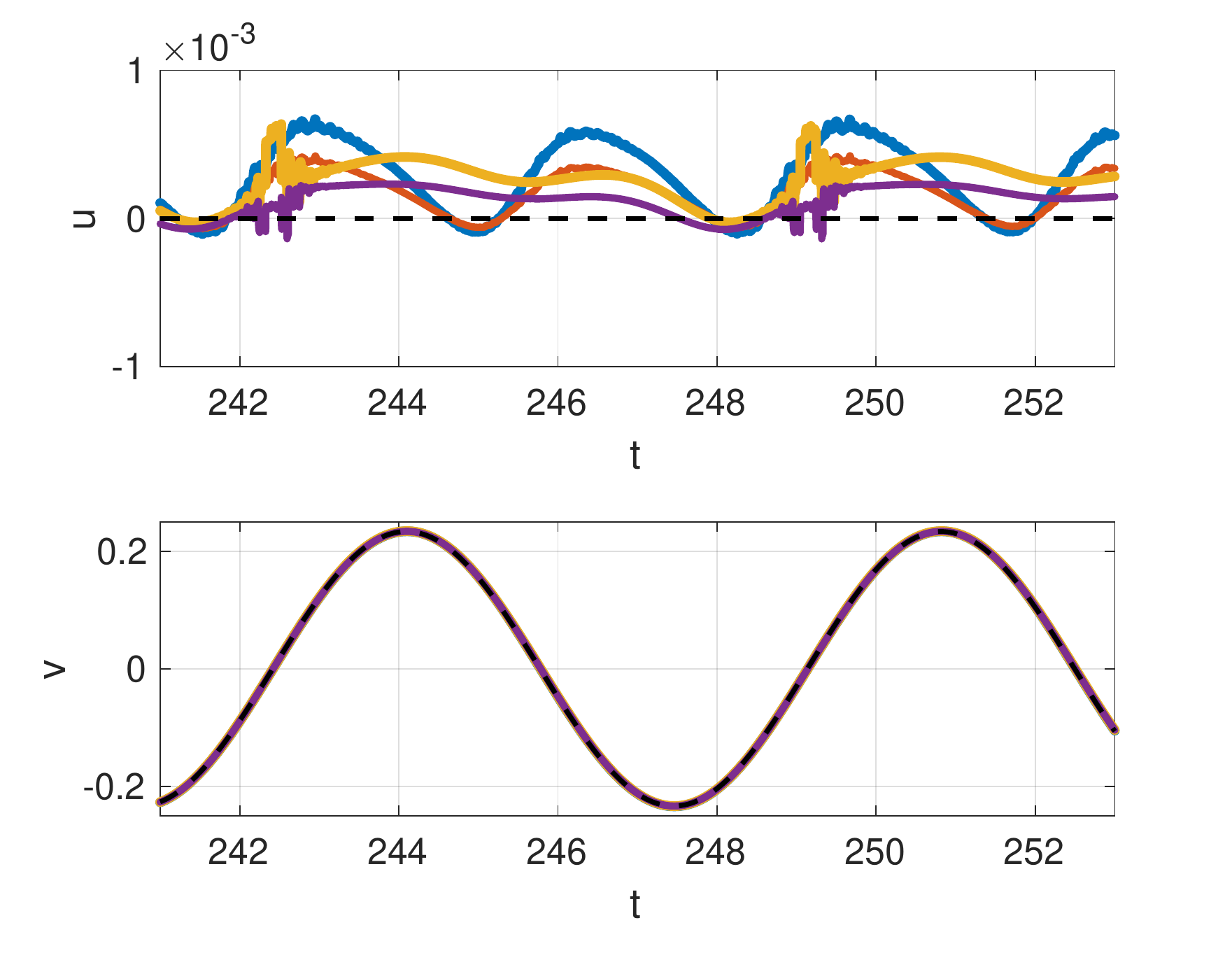}
		\caption{}
	\end{subfigure}
	\centering
	\caption{Moving cylinder: Evolution of velocities at two probe points. Case 1: $P = 2, \IBMparam = 1 \times 10^{-3}$ (volume penalization). Case 2: $P = 2, \IBMparam = 1 \times 10^{-3}, \chi_f = 1 \times 10^{3}, \Delta = 100$ (Combined method). Case 3: $P = 3, \IBMparam = 1 \times 10^{-3}$ (volume penalization). Case 4: $P = 3, \IBMparam = 1 \times 10^{-3}, \chi_f = 1 \times 10^{3}, \Delta = 100$ (Combined method). Reference velocity is shown in the dashed line. a) Point A ($(x, y)/D = (0.2, 0.2)$); b) Point B ($(x, y)/D = (0.41, 0.02)$).}
	\label{fig:Cyl-Flow}
\end{figure*}

To test the accuracy of the numerical scheme, the penalized solution $u_s$ is set to $U$, which varies in space and time to mimic the effect of moving geometry. For the test case, we set the parameters to $A_m = 0, B_m = 1.11, C_m = 0.1, D_m = 10$. The transient solution and error plot is shown in Fig. \ref{fig:MMS-cartoon}. We extend two elements on both sides of the domain $[0,1]$ to mimic the solid element. Note that the number of solid elements in the right region also varies in time, while we ensure at least two elements are penalized in the left solid region. The solution is marched until $t = 0.1$. As shown in Fig. \ref{fig:MMS-cartoon}a, the red-shaded region includes moving boundaries, and is sufficiently large to test the accuracy. In the combined FV+SFD method, the penalization parameter is set to be equal to the time step $\IBMparam = 1 \times 10^{-9}$, with SFD control coefficient $\chi_{f} = 1 \times 10^9$, while the filter width is set to $10^2$. For different polynomial orders and numbers of elements, the error in the L2 norm is computed and compared in \ref{fig:MMS-cartoon}b. As shown in the figure, high-order convergence $N^{-(P+1)}$ is obtained, which is the expected convergence rate of the FR scheme. This indicates that the VP+SFD method does not deteriorate the high-order convergence when the exact solution is imposed inside the moving geometry.

\begin{figure*}[htbp]
	\begin{subfigure}{.48\textwidth}
 		\centering
		\includegraphics[width=0.7\textwidth]{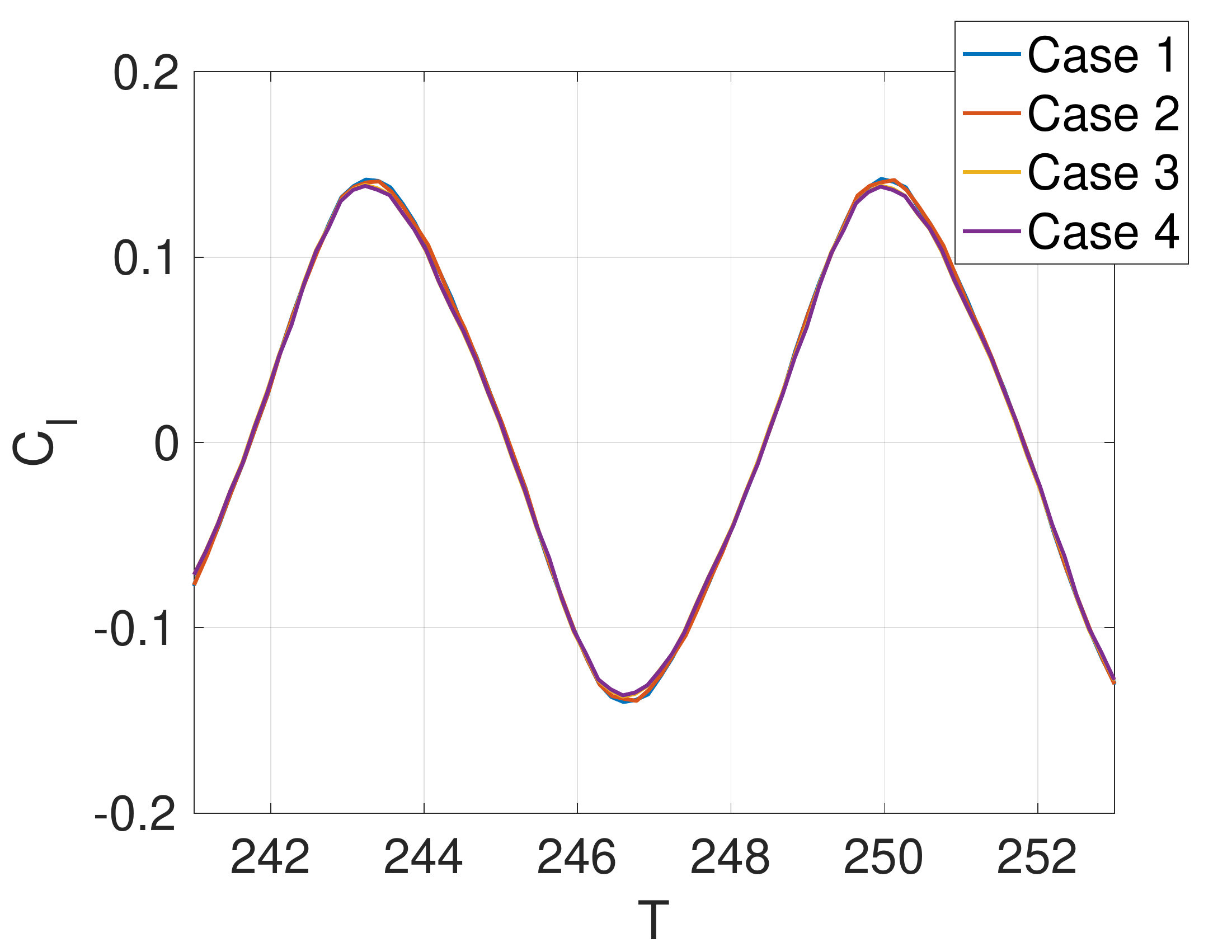}
		\caption{}
	\end{subfigure}
	\begin{subfigure}{.48\textwidth}
 		\centering
		\includegraphics[width=0.7\textwidth]{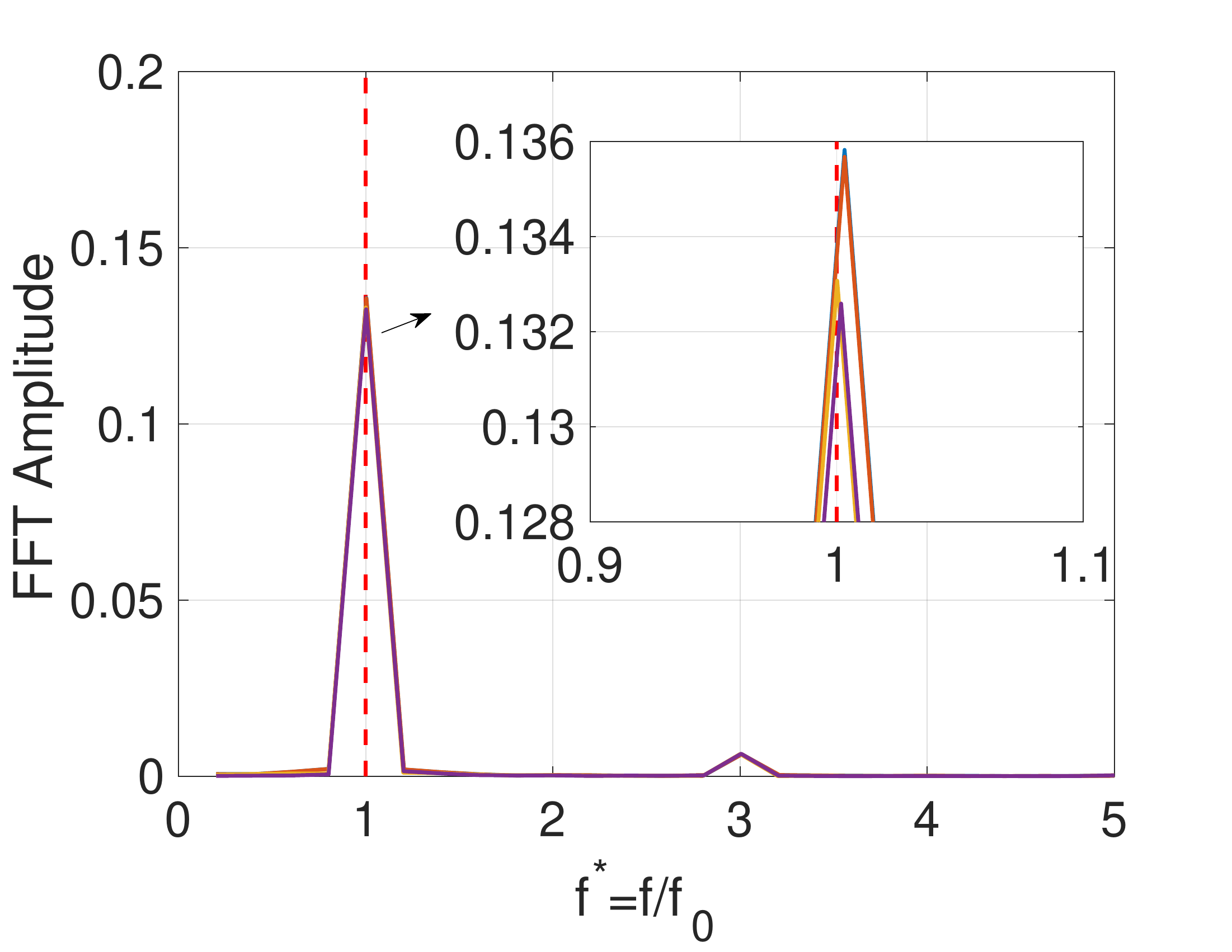}
		\caption{}
	\end{subfigure}
	\centering
	\caption{Moving cylinder: Lift coefficient. a) Time evolution b) Fourier spectrum (vortex shedding frequency is highlighted in the red line).}
	\label{fig:Cyl-Force}
\end{figure*}

\subsection{Flow past a moving circular cylinder}
\label{sec:cylinder}
In the second case, we solve the unsteady flow (i.e., Navier-Stokes) past a moving cylinder at Reynolds number 100 \cite{placzek2009numerical}. The cylinder is free to move in the transverse direction. The background mesh is a rectangular computational domain in $x \in [-30D, 50D]$ and $y \in [-30D, 30D]$, where $D$ is the cylinder diameter. A uniform grid with size $0.03D$ is used in $x \in [-D, D]$ and $y \in [-D, D]$, to maintain accuracy near the cylinder boundary. The total number of elements is $399 \times 228$. Note that the wake region is refined to capture the periodic vortex shedding. For a moving cylinder, the mask function is time-dependent and the non-zero velocities for solid points are imposed according to motion functions. A sinusoidal motion $y(t) = A \text{sin} (2 \pi f_0 t) $ is considered, where $A = 0.25D$. The frequency of harmonic motion is defined as the ratio between forced motion frequency $f_0$ and the vortex shedding frequency $f_s$ (of a static cylinder), where $f_0 = 0.9 f_s$ is focused. As discussed in \cite{placzek2009numerical}, this case exhibits a lock-in phenomenon, where the vortex shedding frequency (obtained by the frequency of lift coefficient, $f$) diverges from the value of static cylinder ($f_s$) at the same Reynolds number, and it locks on the frequency of the forced oscillation ($f / f_0 \approx  1$). Our results show that the vortex shedding frequency $f_s$ is 0.1653 (defined by the Strouhal number), which agrees well with the literature \cite{placzek2009numerical}. We run the simulation until the periodic vortex shedding fully develops and start to move the cylinder. 

The probes are selected as the closest Gauss solution point to two particular coordinates $(x, y)/D = (0.2, 0.2)$ and $(x, y)/D = (0.41, 0.02)$. From Fig. \ref{fig:Cyl-cartoon}, these two points always lie inside the solid element and are approaching the boundary when the minimum plunging displacement is reached. We consider two polynomials $P = 2$ and $P = 3$, and compare the difference between standard VP and the combined VP+SFD method. This leads to four different test cases, as introduced in Fig. \ref{fig:Cyl-Flow}. The time step is set to $1e-4$ and $5e-5$ for polynomial orders $2$ and $3$. To compare the effect of SFD, the penalization parameter is set to a relatively large value, i.e., $\IBMparam = 1 \times 10^{-3}$. 

To evaluate the accuracy in imposing boundary conditions through the IBM, the time evolution of the velocity is compared in Fig. \ref{fig:Cyl-Flow}. As can be clearly seen in the figure, when the combined approach is used, the boundary condition is imposed more accurately, and the oscillation is reduced to approximate the exact velocity. A comparison of the lift coefficient is shown in Fig. \ref{fig:Cyl-Force}, where similar results are produced by all methods. When applying a Fourier transform to the lift coefficient, the varying accuracy in capturing the lift frequency can be studied. As mentioned by Placzek et al. \cite{placzek2009numerical}, due to the lock-in phenomenon, the frequency of the lift coefficient will lock on the vortex shedding frequency. This is well reproduced by the present simulations, where the frequency ratio is always close to one. 

\section{Conclusions}
The recently proposed immersed boundary method combining volume penalization and selective frequency damping has been extended to moving geometries. To do so, the numerical solution inside the solid is decomposed into a predefined movement and an oscillatory part (spurious waves), where the latter is damped by an SFD approach combined with volume penalization. We challenge the method with two cases. First, a new manufactured solution problem is proposed to show that the method can recover high-order accuracy. Second, we validate the methodology by simulating the laminar flow past a moving cylinder, where improved accuracy of the combined method is reported.

\section*{Acknowledgments}
JK would like to thank the support of the Alexander von Humboldt Foundation (Ref 3.5-CHN-1227287-HFST-P). EF would like to thank the support of
Agencia Estatal de Investigación (for the grant "Europa Excelencia 2022" Proyecto EUR2022-134041/AEI/10.13039/501100011033) y del Mecanismo de Recuperación y Resiliencia de la Unión Europea, and the Comunidad de Madrid and Universidad Politécnica de Madrid for the Young Investigators award: APOYO-JOVENES-21-53NYUB-19-RRX1A0. 


\bibliographystyle{model1-num-names}
\bibliography{refs}

\begin{thebibliography}{14}
\expandafter\ifx\csname natexlab\endcsname\relax\def\natexlab#1{#1}\fi
\providecommand{\url}[1]{\texttt{#1}}
\providecommand{\href}[2]{#2}
\providecommand{\path}[1]{#1}
\providecommand{\DOIprefix}{doi:}
\providecommand{\ArXivprefix}{arXiv:}
\providecommand{\URLprefix}{URL: }
\providecommand{\Pubmedprefix}{pmid:}
\providecommand{\doi}[1]{\href{http://dx.doi.org/#1}{\path{#1}}}
\providecommand{\Pubmed}[1]{\href{pmid:#1}{\path{#1}}}
\providecommand{\bibinfo}[2]{#2}
\ifx\xfnm\relax \def\xfnm[#1]{\unskip,\space#1}\fi
\bibitem[{Wang et~al.(2013)Wang, Fidkowski, Abgrall, Bassi, Caraeni, Cary,
  Deconinck, Hartmann, Hillewaert, Huynh et~al.}]{wang2013high}
\bibinfo{author}{Z.~J. Wang}, \bibinfo{author}{K.~Fidkowski},
  \bibinfo{author}{R.~Abgrall}, \bibinfo{author}{F.~Bassi},
  \bibinfo{author}{D.~Caraeni}, \bibinfo{author}{A.~Cary},
  \bibinfo{author}{H.~Deconinck}, \bibinfo{author}{R.~Hartmann},
  \bibinfo{author}{K.~Hillewaert}, \bibinfo{author}{H.~T. Huynh}, et~al.,
\newblock \bibinfo{title}{High-order cfd methods: current status and
  perspective},
\newblock \bibinfo{journal}{International Journal for Numerical Methods in
  Fluids} \bibinfo{volume}{72} (\bibinfo{year}{2013})
  \bibinfo{pages}{811--845}.
\bibitem[{Mittal and Iaccarino(2005)}]{mittal2005immersed}
\bibinfo{author}{R.~Mittal}, \bibinfo{author}{G.~Iaccarino},
\newblock \bibinfo{title}{Immersed boundary methods},
\newblock \bibinfo{journal}{Annu. Rev. Fluid Mech.} \bibinfo{volume}{37}
  (\bibinfo{year}{2005}) \bibinfo{pages}{239--261}.
\bibitem[{Sotiropoulos and Yang(2014)}]{sotiropoulos2014immersed}
\bibinfo{author}{F.~Sotiropoulos}, \bibinfo{author}{X.~Yang},
\newblock \bibinfo{title}{Immersed boundary methods for simulating
  fluid-structure interaction},
\newblock \bibinfo{journal}{Progress in Aerospace Sciences}
  \bibinfo{volume}{65} (\bibinfo{year}{2014}) \bibinfo{pages}{1--21}.
\bibitem[{Griffith and Patankar(2020)}]{griffith2020immersed}
\bibinfo{author}{B.~E. Griffith}, \bibinfo{author}{N.~A. Patankar},
\newblock \bibinfo{title}{Immersed methods for fluid--structure interaction},
\newblock \bibinfo{journal}{Annual Review of Fluid Mechanics}
  \bibinfo{volume}{52} (\bibinfo{year}{2020}) \bibinfo{pages}{421--448}.
\bibitem[{Fidkowski and Darmofal(2007)}]{fidkowski2007triangular}
\bibinfo{author}{K.~J. Fidkowski}, \bibinfo{author}{D.~L. Darmofal},
\newblock \bibinfo{title}{A triangular cut-cell adaptive method for high-order
  discretizations of the compressible navier--stokes equations},
\newblock \bibinfo{journal}{Journal of Computational Physics}
  \bibinfo{volume}{225} (\bibinfo{year}{2007}) \bibinfo{pages}{1653--1672}.
\bibitem[{M{\"u}ller et~al.(2017)M{\"u}ller, Kr{\"a}mer-Eis, Kummer, and
  Oberlack}]{muller2017high}
\bibinfo{author}{B.~M{\"u}ller}, \bibinfo{author}{S.~Kr{\"a}mer-Eis},
  \bibinfo{author}{F.~Kummer}, \bibinfo{author}{M.~Oberlack},
\newblock \bibinfo{title}{A high-order discontinuous galerkin method for
  compressible flows with immersed boundaries},
\newblock \bibinfo{journal}{International Journal for Numerical Methods in
  Engineering} \bibinfo{volume}{110} (\bibinfo{year}{2017})
  \bibinfo{pages}{3--30}.
\bibitem[{Kou et~al.(2022{\natexlab{a}})Kou, Joshi, Hurtado-de Mendoza, Puri,
  Hirsch, and Ferrer}]{kou2021IBMFR2}
\bibinfo{author}{J.~Kou}, \bibinfo{author}{S.~Joshi},
  \bibinfo{author}{A.~Hurtado-de Mendoza}, \bibinfo{author}{K.~Puri},
  \bibinfo{author}{C.~Hirsch}, \bibinfo{author}{E.~Ferrer},
\newblock \bibinfo{title}{Immersed boundary method for high-order flux
  reconstruction based on volume penalization},
\newblock \bibinfo{journal}{Journal of Computational Physics}
  \bibinfo{volume}{448} (\bibinfo{year}{2022}{\natexlab{a}})
  \bibinfo{pages}{110721}.
\bibitem[{Kou et~al.(2022{\natexlab{b}})Kou, Hurtado-de Mendoza, Joshi,
  Le~Clainche, and Ferrer}]{kou2021VonNeumann}
\bibinfo{author}{J.~Kou}, \bibinfo{author}{A.~Hurtado-de Mendoza},
  \bibinfo{author}{S.~Joshi}, \bibinfo{author}{S.~Le~Clainche},
  \bibinfo{author}{E.~Ferrer},
\newblock \bibinfo{title}{Eigensolution analysis of immersed boundary method
  based on volume penalization: Applications to high-order schemes},
\newblock \bibinfo{journal}{Journal of Computational Physics}
  \bibinfo{volume}{449} (\bibinfo{year}{2022}{\natexlab{b}})
  \bibinfo{pages}{110817}.
\bibitem[{Kou and Ferrer(2023)}]{kou2022combined}
\bibinfo{author}{J.~Kou}, \bibinfo{author}{E.~Ferrer},
\newblock \bibinfo{title}{A combined volume penalization/selective frequency
  damping approach for immersed boundary methods applied to high-order
  schemes},
\newblock \bibinfo{journal}{Journal of Computational Physics}
  (\bibinfo{year}{2023}) \bibinfo{pages}{111678}.
\bibitem[{Angot et~al.(1999)Angot, Bruneau, and Fabrie}]{angot1999penalization}
\bibinfo{author}{P.~Angot}, \bibinfo{author}{C.-H. Bruneau},
  \bibinfo{author}{P.~Fabrie},
\newblock \bibinfo{title}{A penalization method to take into account obstacles
  in incompressible viscous flows},
\newblock \bibinfo{journal}{Numerische Mathematik} \bibinfo{volume}{81}
  (\bibinfo{year}{1999}) \bibinfo{pages}{497--520}.
\bibitem[{Schneider(2015)}]{schneider2015immersed}
\bibinfo{author}{K.~Schneider},
\newblock \bibinfo{title}{Immersed boundary methods for numerical simulation of
  confined fluid and plasma turbulence in complex geometries: a review},
\newblock \bibinfo{journal}{Journal of Plasma Physics} \bibinfo{volume}{81}
  (\bibinfo{year}{2015}) \bibinfo{pages}{435810601}.
\bibitem[{{\AA}kervik et~al.(2006){\AA}kervik, Brandt, Henningson,
  H{\oe}pffner, Marxen, and Schlatter}]{aakervik2006steady}
\bibinfo{author}{E.~{\AA}kervik}, \bibinfo{author}{L.~Brandt},
  \bibinfo{author}{D.~S. Henningson}, \bibinfo{author}{J.~H{\oe}pffner},
  \bibinfo{author}{O.~Marxen}, \bibinfo{author}{P.~Schlatter},
\newblock \bibinfo{title}{Steady solutions of the navier-stokes equations by
  selective frequency damping},
\newblock \bibinfo{journal}{Physics of fluids} \bibinfo{volume}{18}
  (\bibinfo{year}{2006}) \bibinfo{pages}{068102}.
\bibitem[{Tremblay et~al.(2006)Tremblay, Etienne, and
  Pelletier}]{tremblay2006code}
\bibinfo{author}{D.~Tremblay}, \bibinfo{author}{S.~Etienne},
  \bibinfo{author}{D.~Pelletier},
\newblock \bibinfo{title}{Code verification and the method of manufactured
  solutions for fluid-structure interaction problems},
\newblock in: \bibinfo{booktitle}{36th AIAA Fluid Dynamics Conference and
  Exhibit}, \bibinfo{year}{2006}, p. \bibinfo{pages}{3218}.
\bibitem[{Placzek et~al.(2009)Placzek, Sigrist, and
  Hamdouni}]{placzek2009numerical}
\bibinfo{author}{A.~Placzek}, \bibinfo{author}{J.-F. Sigrist},
  \bibinfo{author}{A.~Hamdouni},
\newblock \bibinfo{title}{Numerical simulation of an oscillating cylinder in a
  cross-flow at low reynolds number: Forced and free oscillations},
\newblock \bibinfo{journal}{Computers \& Fluids} \bibinfo{volume}{38}
  (\bibinfo{year}{2009}) \bibinfo{pages}{80--100}.

\end{thebibliography}

\end{document}